\documentclass[preprint2]{aastex}

\shorttitle{BEST II Catalog of Variable Stars: CoRoT SRc02 field}
\shortauthors{Klagyivik et al.}

\begin{document}

\title{The Berlin Exoplanet Search Telescope II Catalog of Variable Stars. II. Characterization of the CoRoT SRc02 field\\
}

\author{P.~Klagyivik\altaffilmark{1,2,3},
Sz.~Csizmadia\altaffilmark{3},
T.~Pasternacki\altaffilmark{3},
J.~Cabrera\altaffilmark{3},
R.~Chini\altaffilmark{4,5},
P.~Eigm\"uller\altaffilmark{3},
A.~Erikson\altaffilmark{3},
T.~Fruth\altaffilmark{3,6},
P.~Kabath\altaffilmark{7},
R.~Lemke\altaffilmark{4},
M.~Murphy\altaffilmark{8},
H.~Rauer\altaffilmark{3,9}, and
R.~Titz-Weider\altaffilmark{3}}

\altaffiltext{1}{Instituto de Astrof\'isica de Canarias, E-38200 La Laguna, Tenerife, Spain}
\altaffiltext{2}{Universidad de La Laguna, Dept. Astrof\'isica, E-38206 La Laguna, Tenerife, Spain}
\altaffiltext{3}{Institut f\"ur Planetenforschung, Deutsches Zentrum f\"ur Luft- und Raumfahrt, Rutherfordstra\ss e 2, 12489 Berlin, Germany}
\altaffiltext{4}{Astronomisches Institut, Ruhr-Universit\"at Bochum, 44780 Bochum, Germany}
\altaffiltext{5}{Instituto de Astronom\'ia, Universidad Cat\'olica del Norte, Antofagasta, Chile}
\altaffiltext{6}{German Space Operations Center, German Aerospace Center, M\"unchener Strasse 20, 82234 We{\ss}ling, Germany}
\altaffiltext{7}{Astronomical Institute of The Czech Academy of Sciences, Fri\v{c}ova 298, 25165, Ond\v{r}ejov, Czech Republic}
\altaffiltext{8}{Depto. F\'isica, Universidad Cat\'olica del Norte, PO 1280, Antofagasta, Chile}
\altaffiltext{9}{Technische Universit\"at Berlin, Zentrum f\"ur Astronomie und Astrophysik, Hardenbergstra\ss e 36, 10623 Berlin, Germany}

\begin{abstract}
Time-series photometry of the CoRoT field SRc02 was obtained by the
Berlin Exoplanet Search Telescope II (BEST II) in 2009.
The main aim was the ground based follow-up of the CoRoT field in order
to detect variable stars with better spatial resolution than
what can be achieved with the CoRoT space telescope.
A total of 1,846 variable stars were detected,
of which only 30 have been previously known.
For nine eclipsing binaries the stellar parameters were determined
by modeling their light curve.
\end{abstract}


\section{Introduction}
\label{intro}

CoRoT has been a 27 cm diameter space telescope equipped with four CCD-cameras.
Two of them were used for observing a dozen of pulsational and other kinds of
light variations of bright stars with high time-sampling (so-called asteroseismological channel),
while other two CCDs were used to search for transit signals of about 6000 stars
per CCD (so-called exo-channel). In the exo-channel there were biprisms in front
of the CCDs which produced a low ($\Delta\lambda / \lambda \approx 3$) resolution spectra
for each stars. This avoided the saturation of many stars, and like defocusing,
helped to increase the signal-to-nois ratio. Because of available telemetry rate,
most of the photometric masks of the pre-selected stars were read-out in
every 512 seconds, but a few hundred stars had a flux measurement in every
32 seconds as well as for a few hundred stars different parts of the photometric
masks were read-out separately, yielding the so-called CoRoT-blue, -green and -red
light curves. For details see \citet{baglin07} and \citet{auvergne09}.

But due to this technique, the point spread function of CoRoT's exo-channel
was typically 80$\times$20 arcseconds and was varying from star to star.
Therefore nearby stars in the frame can pollute the observed stars,
spreading a certain amount of their flux inside the photometric mask.
Such a polluting star is called contaminating star in CoRoT-terminology.
The process how to determine which star is contaminating and how much is
described in \citet{pasternacki11a} and \citet{gardes11}, for instance.
However, in those calculations the contaminating star is considered as a constant star.
If it is variable then it is possible that a contaminating source is a foreground
or background eclipsing binary and small amount of its light is contaminating
the main target, causing a false positive as a diluted, small transit-like signal
in the CoRoT-target. Therefore additional, ground-based, higher spatial resolution
photometry is necessary to filter out such cases. \citet{deeg09} describes
how different kinds of such photometric studies help to decide about the true
nature of the observed CoRoT transit signals. Our previous
\citep{karo07, kaba07, kaba08, kaba09a, kaba09b, klagyivik13, fruth12, fruth13} and present works
reports our contribution to this as well as reports the detected variables and
their basic properties. Future studies on CoRoT light curves can utilize these
data to remove the variability stemming from a possible contaminator.

Beyond the much better angular resolution the main advantage of our data is the different
epoch of the observation. For periodic sources (e.g. eclipsing binaries, planetary candidates)
this helps to determine a more accurate period, while for non-periodic sources a much longer
observation is abailable, which is helpful to describe the nature of the light variation.

In this paper we present our independent study on the variable stars
in the direction of the CoRoT field SRc02 detected by BEST II.
Section~\ref{data} presents the observations and the description of the telescope used.
In Section~\ref{variable_stars} we present the variable star selection method
and the description of the classification scheme. The previously known
and the newly detected variable stars are in Section~\ref{firstresults}.
In Section~\ref{binaries} we present the fitted models of nine selected eclipsing binaries.
Eventually the summary and conclusions of this paper can be found in Section~\ref{summary}.

\section{Observations and data reduction}
\label{data}

The observations were performed with the BEST II telescope located
at the Uni\-versi\-t\"ats\-stern\-war\-te Bochum near Observatorio Cerro Armazones, Chile.
The system consists of a Takahashi 25\,cm Baker-Ritchey-Chr\'etien telescope equipped with
a 4k $\times$ 4k Finger Lakes CCD. The corresponding field of view is
$1.7^\circ\times 1.7^\circ$ with an angular resolution of $1\farcs5$ pixel$^{-1}$.
In order to maximize the photon yield and to get more accurate photometry
of the fainter stars no filter was used. The exposure time
was 120\,s for all the images.

\begin{figure*}
\centering
\includegraphics[width=0.95\textwidth]{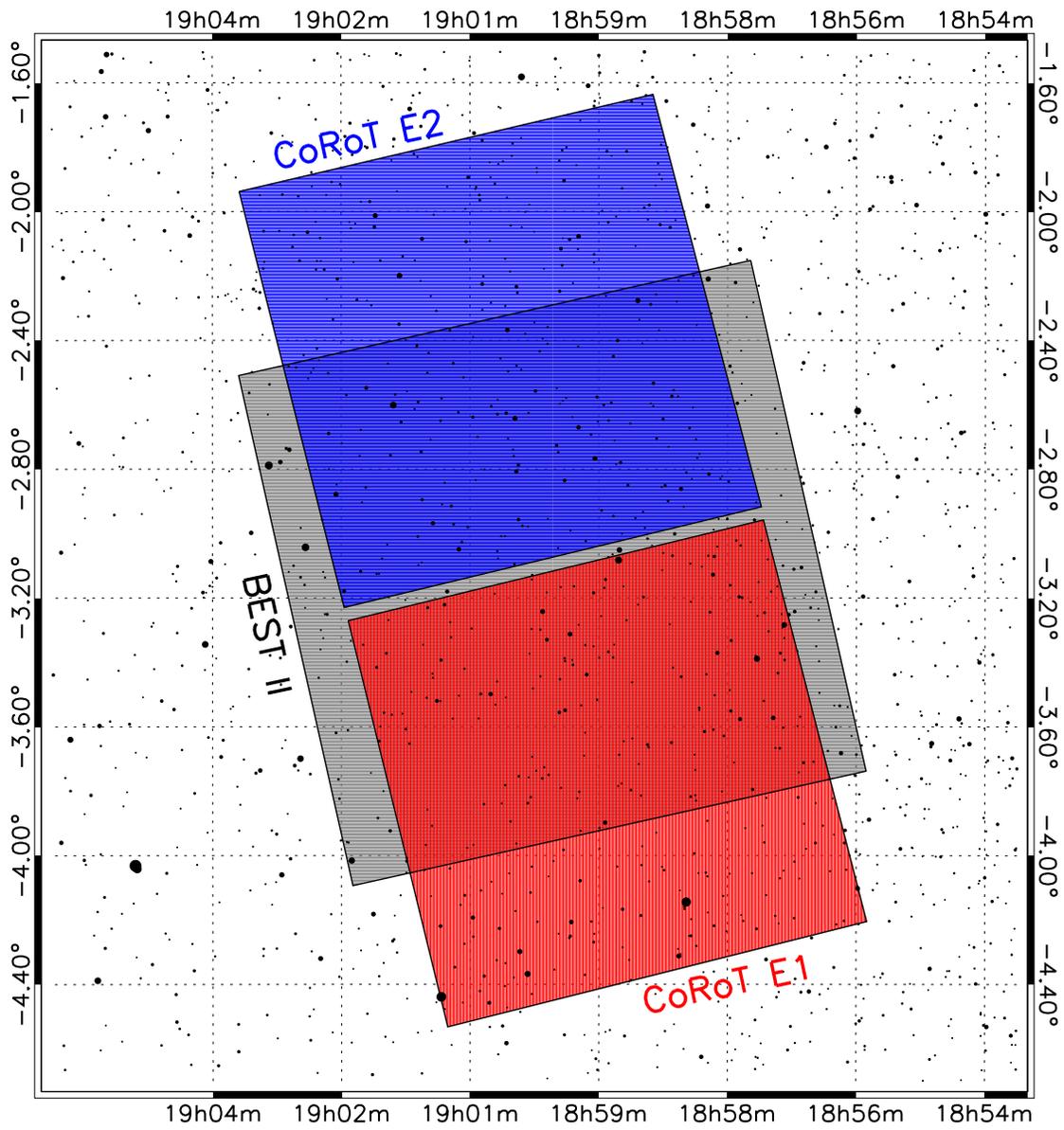}
\caption{Comparison of the fields observed by CoRoT and BEST II.}
\label{fig:field}
\end{figure*}

BEST II observed the CoRoT target field SRc02 during a total of
32 nights between 2009 May 4 and 2009 July 28.
An illustration of the BEST II and the CoRoT field SRc02 is shown in Figure \ref{fig:field}.
$\sim65\%$ of the CoRoT field is covered by BEST II, which was centered on the coordinates:
\begin{eqnarray*}
 \alpha(J2000.0) & = & 18^{\mathrm h}\,59^{\mathrm m}\,47^{\mathrm s} \\
 \delta(J2000.0) & = & -03\degr\,07\arcmin\,37\arcsec.
\end{eqnarray*}

The acquired observations were processed using the BEST automated
photometric pipeline as described in \citet{kaba09a}, \citet{raue10}, \citet{pasternacki11b} and \citet{fruth12}.
The resulting datasets consist of 1,307 observations of 86,944 stellar objects.
Note that CoRoT observed 11,408 targets in SRc02 field, which is much less than the targets
observed by BEST II. The most important difference is the target selection. Instead of observing all targets down to
a certain magnitude limit, CoRoT observed a pre-selected sample of stars optimized for tranziting planet detection
and stellar pulsation studies. BEST II has much better angular resolution than CoRoT, which has a point spread function
of $\sim80\farcs0 \times 20\farcs0$, however, the typical number of BEST II objects in a CoRoT
PSF is only 1-3. Another important difference is the limiting magnitude which is $\sim$2-3 magnitudes deeper for our current survey.

\begin{figure*}
\centering
\includegraphics[width=0.95\textwidth]{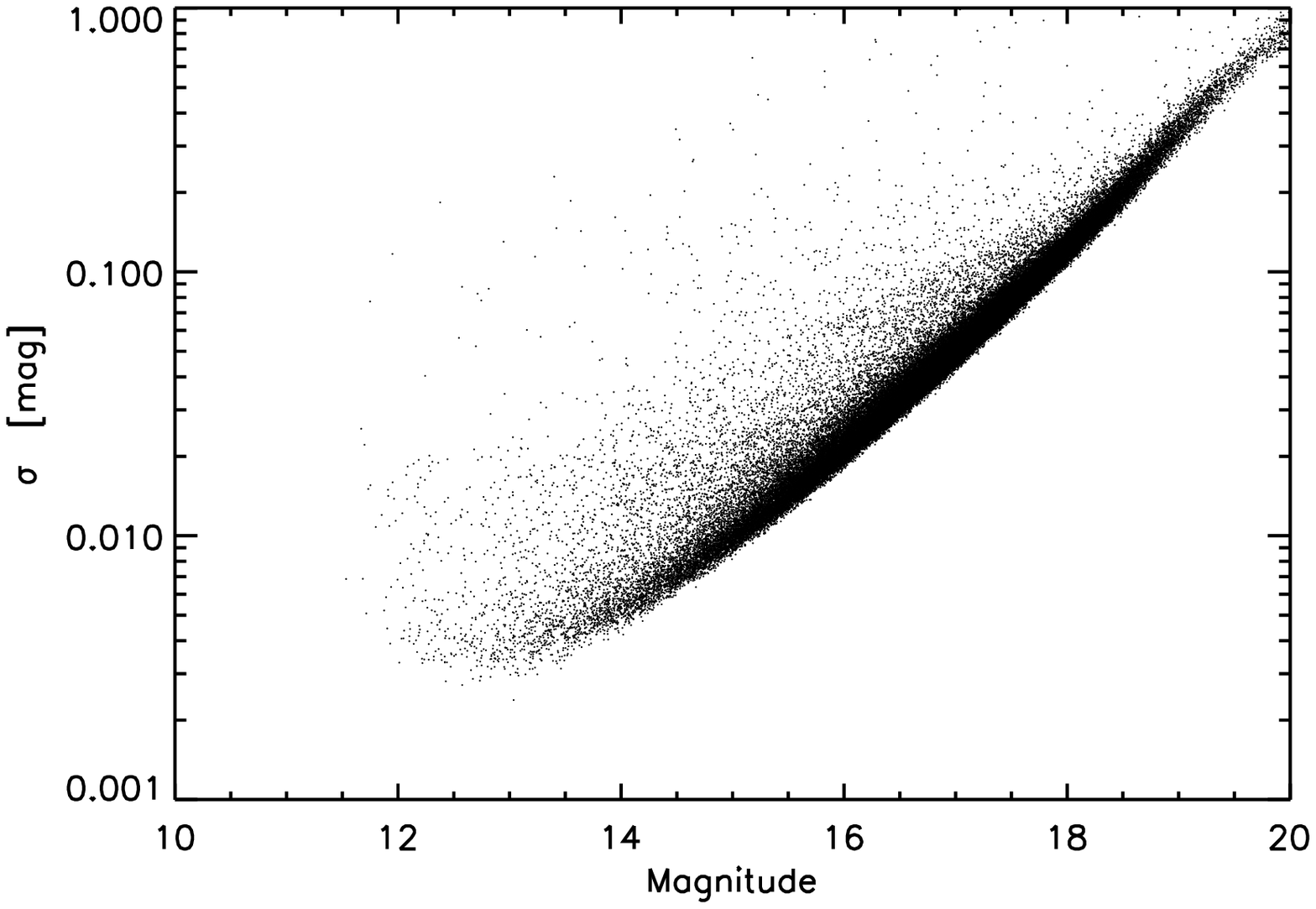}
\caption{Photometric accuracy of the detected sources.}
\label{fig:phot_acc}
\end{figure*}

All stars are matched with the UCAC3 catalog (Zacharias et al. 2010) in order
to assign equatorial coordinates and to adjust instrumental magnitudes to a standard magnitude system.
The astrometric calibration achieves an average residual of 0.23 arcsecond.
The magnitude calibration is obtained by shifting each data set by the
median difference between all instrumental magnitudes and their respective
catalog value (R2MAG of UCAC3). Since the photometric systems are comparable
but not identical, this calibration yields an absolute accuracy of $\sim$0.5\,mag
and ranges from 12 to 20\,mag. The number of stars measured
below 1\% relative accuracy is 4,535. The relative photometric accuracy
of all targets are shown in Figure \ref{fig:phot_acc}.

\section{Variable stars}
\label{variable_stars}

\subsection{Detection}
\label{detection}

For detecting the variable sources, we apply the well functioning method
described by \citet{fruth12}. It is based on the widely-used variability
index $J$ \citep{stet96,zhan03} and a multiharmonic period search
\citep{sccz96}, but also involves an automatic process
of dealing with systematic variability. All light curves with $J > 0$ ($86\%$)
were fitted with seven harmonics and ranked using the modified
Analysis-of-Variance statistic $q$ \citep[see][]{fruth12}. A cut-off
limit was set to $q > 5$ based on empirical experiments, resulting in
7,677 variable star candidates.

All candidates were inspected visually and classified on an
individual basis. We detected a total of 1,846 variable stars,
of which only 30 were previously known and 1,816 are new discoveries.

\begin{figure*}
\centering
\includegraphics[width=0.24\textwidth]{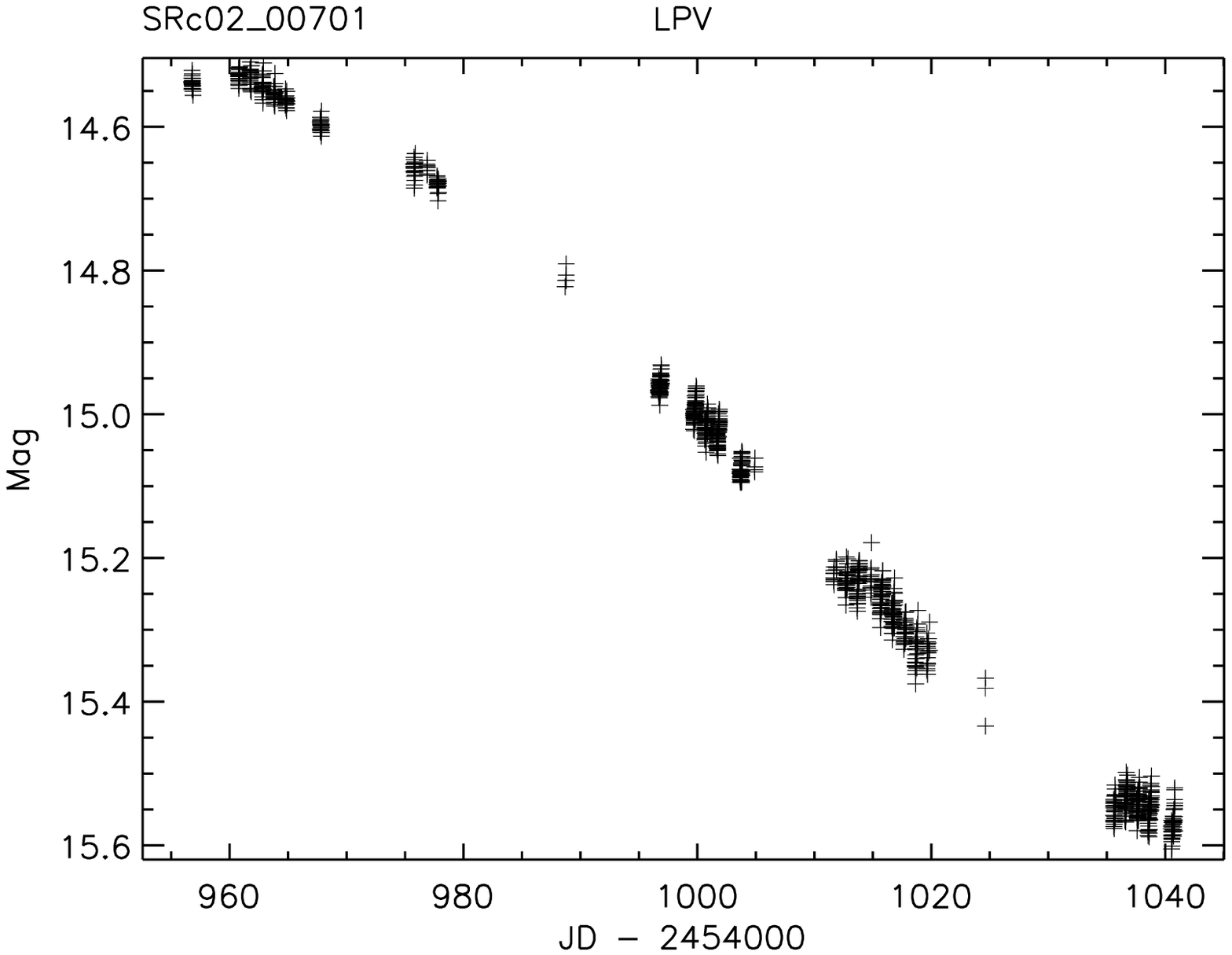}
\includegraphics[width=0.24\textwidth]{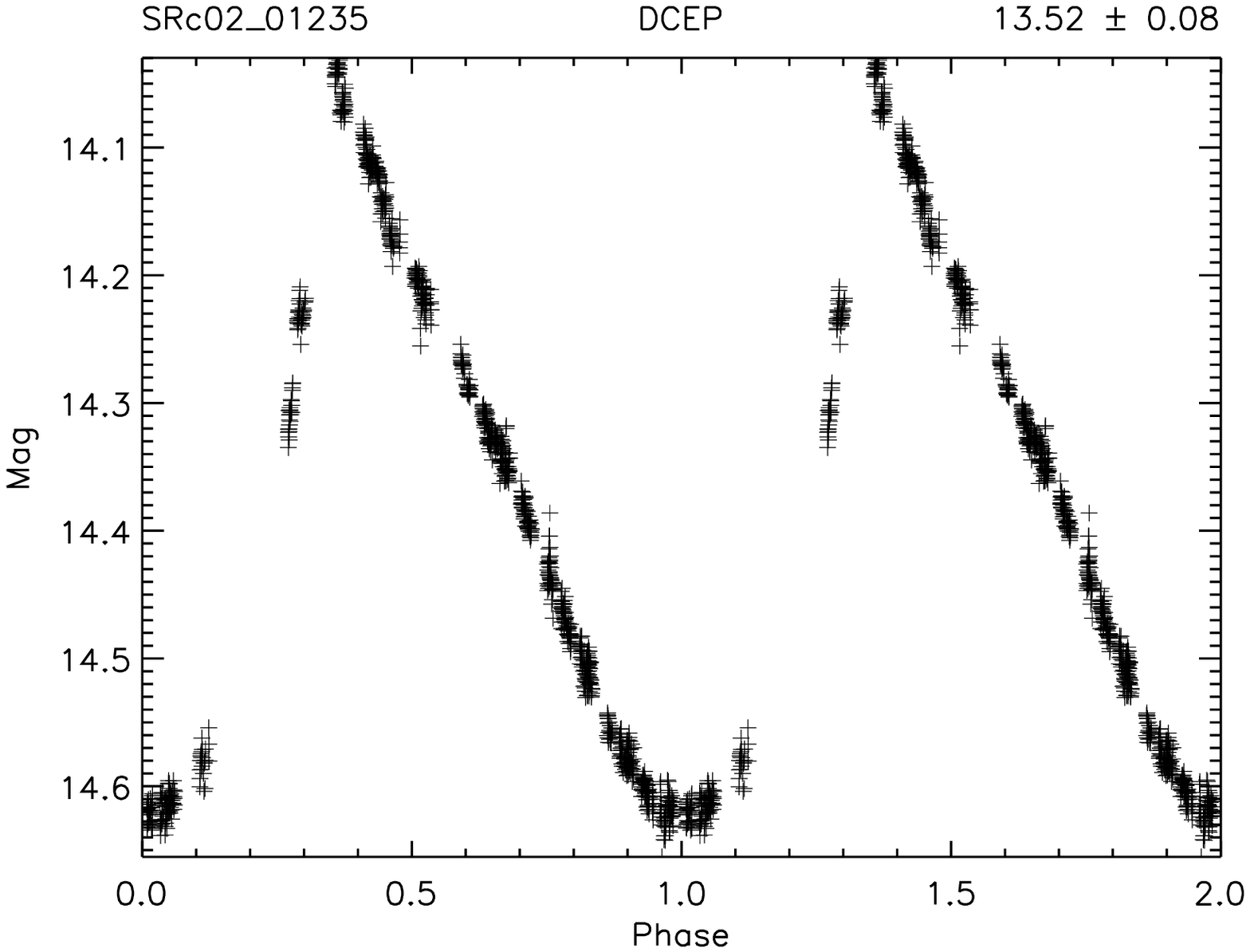}
\includegraphics[width=0.24\textwidth]{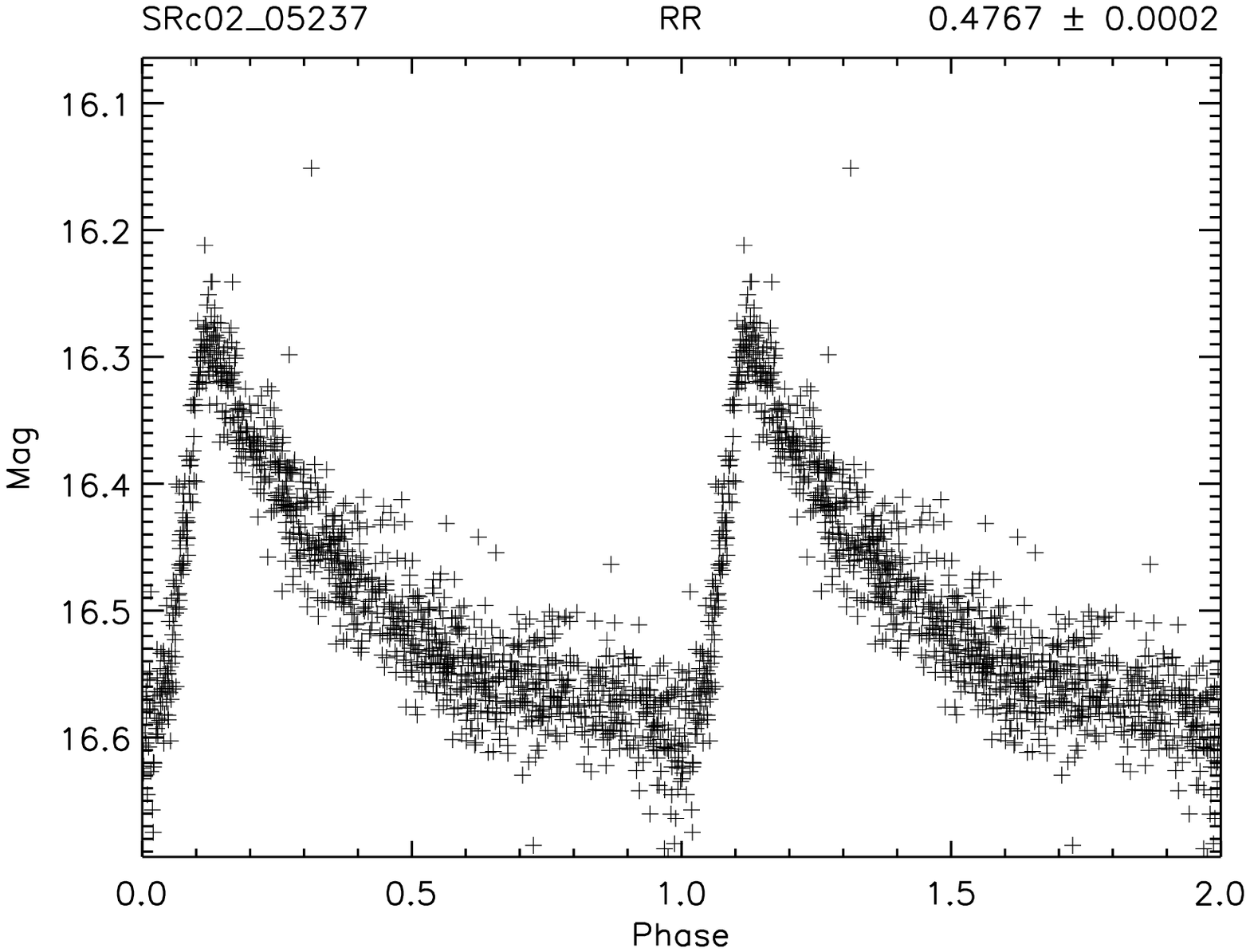}
\includegraphics[width=0.24\textwidth]{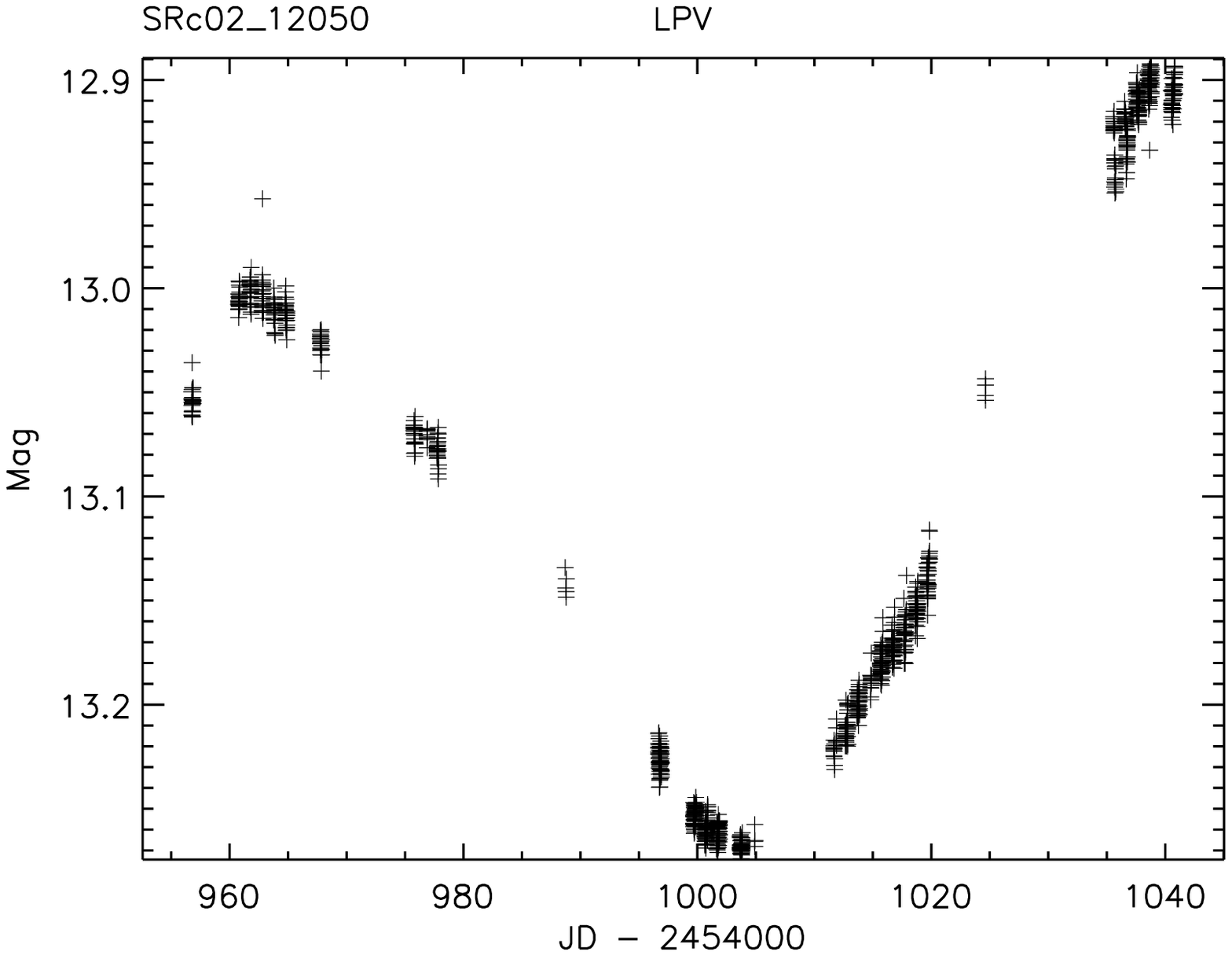}
\includegraphics[width=0.24\textwidth]{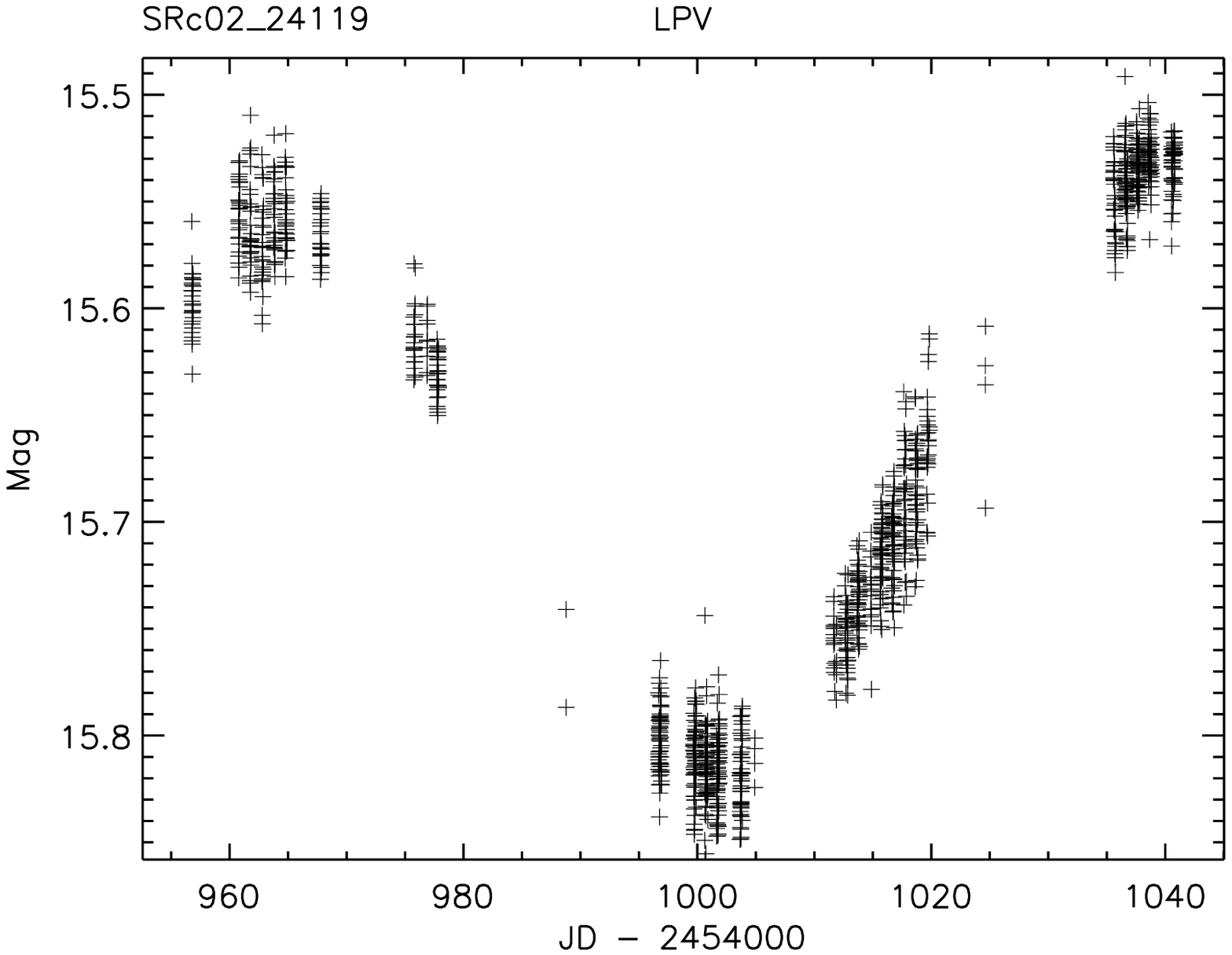}
\includegraphics[width=0.24\textwidth]{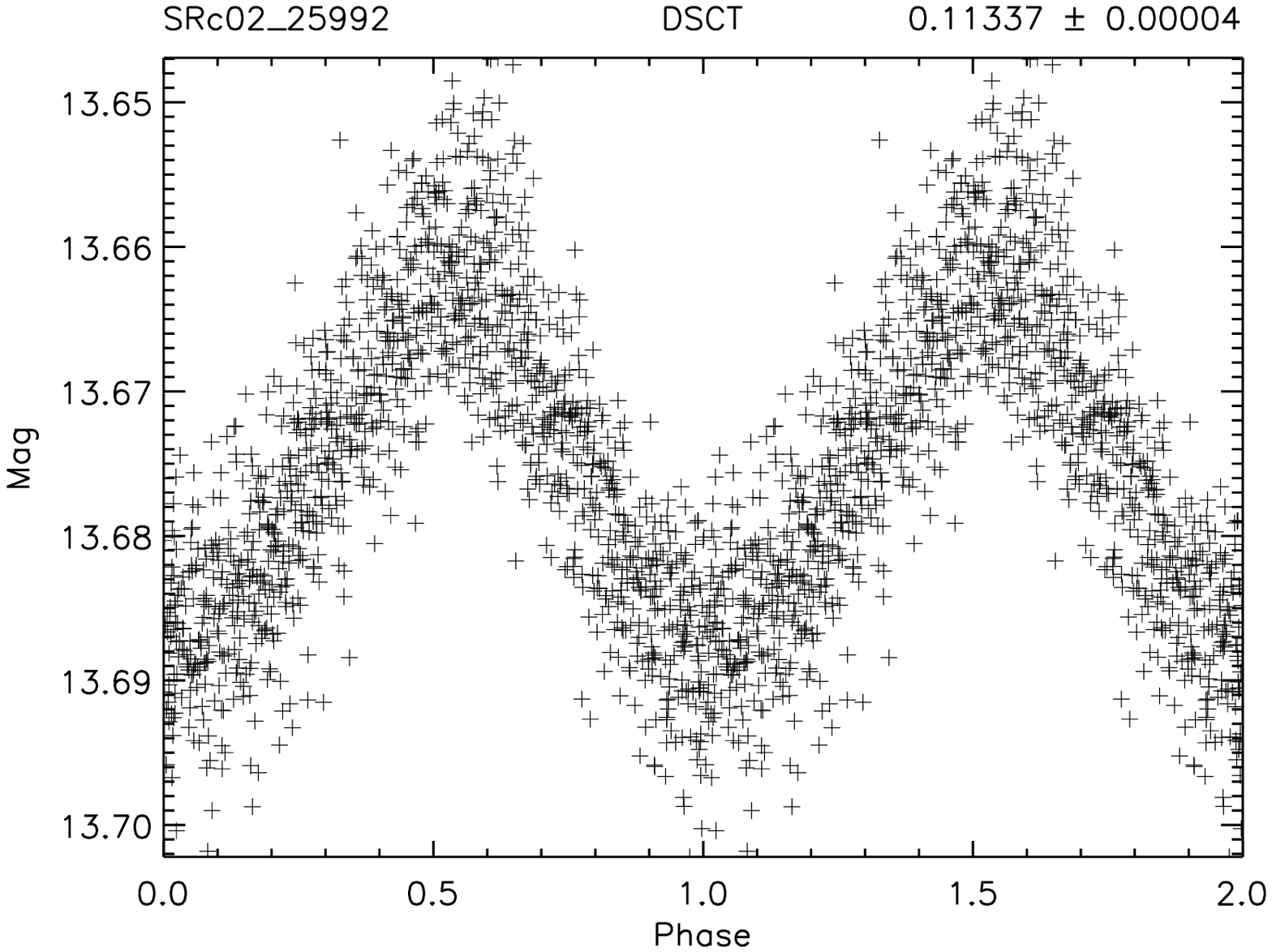}
\includegraphics[width=0.24\textwidth]{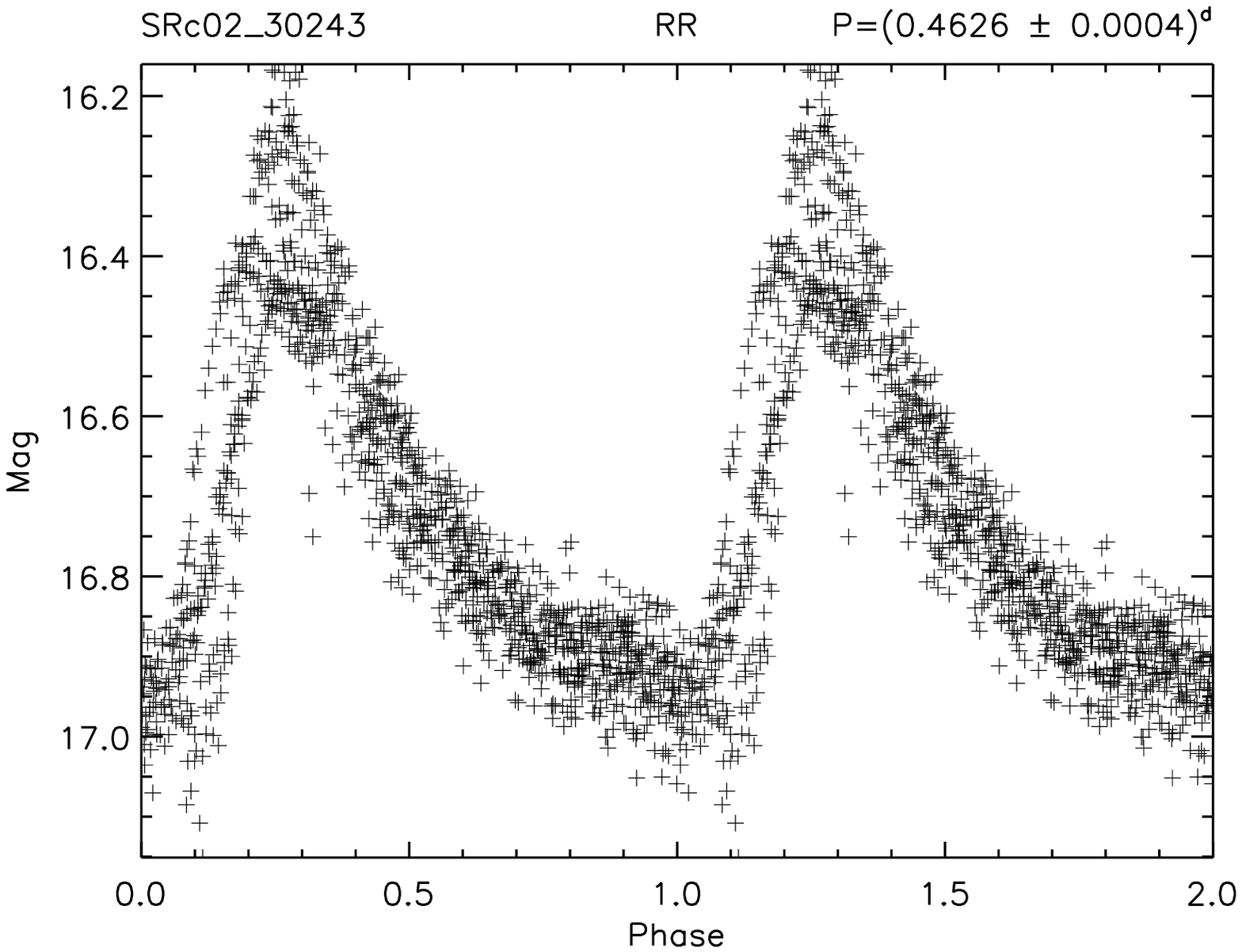}
\includegraphics[width=0.24\textwidth]{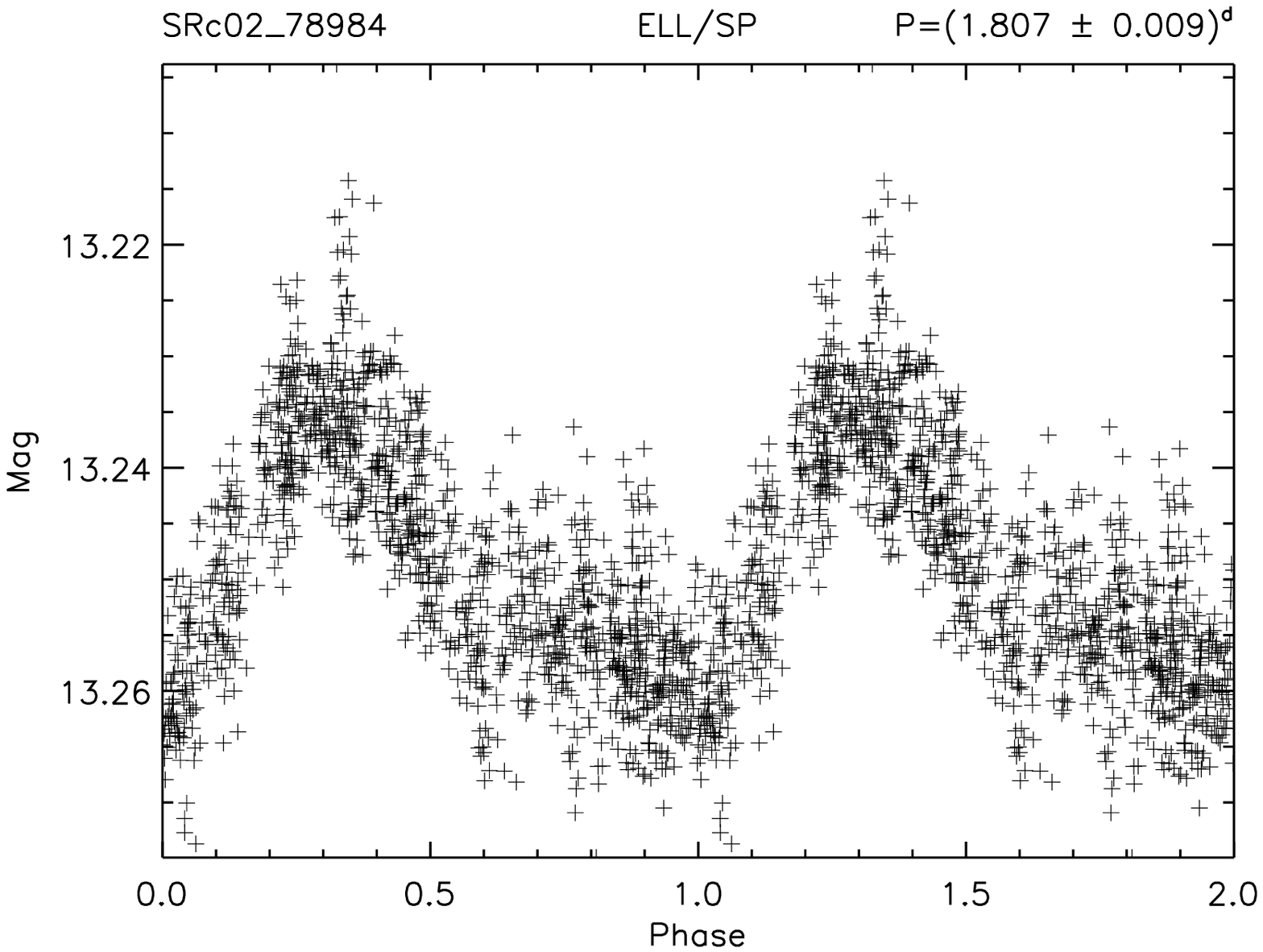}
\caption{Sample light curves of some selected variable stars. For periodic stars the phase folded light curve is shown.
All the light curves are available in the electronic edition of the journal.}
\label{fig:lc_sample}
\end{figure*}

The variable star catalog and the observed light curves are presented here in Table \ref{tab:varcat}
and Figure \ref{fig:lc_sample}, respectively, for guidance regarding its form and content.
Table \ref{tab:varcat} and Figure \ref{fig:lc_sample} are published in its
entirety in the electronic edition of {\it Astronomical Journal}.
The data are available upon request from the co-author A. Erikson (\mbox{anders.erikson@dlr.de}).

\begin{deluxetable}{lcccccccccc}
\rotate
\tablecolumns{11}
\tablewidth{0pc}
\tabletypesize{\scriptsize} 
\tablecaption{\scriptsize Catalog of variable stars detected in {\it CoRoT} field SRc02, sorted by internal BEST II identifiers. \label{tab:varcat}}
\tablehead{
\colhead{BEST ID} & \colhead{flag} & \colhead{2MASS ID} & \colhead{$\alpha(J2000.0)$} & \colhead{$\delta(J2000.0)$} & 
\colhead{R$_B$ [mag]} &\colhead{T$_0$ [rHJD]} & \colhead{P [d]} & \colhead{A [mag]} &  \colhead{Type} & \colhead{Other names}
}
\startdata
SRc02\_00036  &    &18564181-0249366  &$18^h56^m41.8^s$  &$-02^\circ49'36.0"$  &14.63  &  \nodata  &       \nodata             &         \nodata      &LPV  &\\
SRc02\_00049  &c   &18561019-0324329  &$18^h56^m10.2^s$  &$-03^\circ24'33.1"$  &14.98  &   60.721  &      $0.9233 \pm 0.0003$  &          $0.69 \pm 0.02$  &EA  &\\
SRc02\_00148  &    &18562223-0311543  &$18^h56^m22.2^s$  &$-03^\circ11'54.3"$  &14.91  &  \nodata  &       \nodata             &         \nodata      &MISC  &\\
SRc02\_00278  &    &18555391-0344030  &$18^h55^m53.9^s$  &$-03^\circ44'03.1"$  &14.92  &   61.977  &          $7.53 \pm 0.05$  &          $0.14 \pm 0.05$  &EA  &\\
SRc02\_00309  &c   &18560034-0337067  &$18^h56^m00.3^s$  &$-03^\circ37'06.8"$  &14.57  &  \nodata  &       \nodata             &         \nodata      &MISC  &\\
SRc02\_00371  &    &18564095-0252272  &$18^h56^m40.9^s$  &$-02^\circ52'27.1"$  &15.77  &   56.923  &      $0.1832 \pm 0.0002$  &          $0.04 \pm 0.02$  &DSCT  &\\
SRc02\_00427  &    &18560426-0333164  &$18^h56^m04.3^s$  &$-03^\circ33'16.4"$  &15.09  &   60.726  &        $0.821 \pm 0.002$  &          $0.05 \pm 0.02$  &RR  &\\
SRc02\_00436  &    &18570987-0221013  &$18^h57^m09.9^s$  &$-02^\circ21'01.6"$  &15.74  &   56.852  &    $0.09014 \pm 0.00005$  &          $0.03 \pm 0.03$  &DSCT  &\\
SRc02\_00438  &    &18564549-0247448  &$18^h56^m45.5^s$  &$-02^\circ47'44.5"$  &14.65  &  \nodata  &       \nodata             &         \nodata      &MISC  &\\
SRc02\_00458  &    &18565190-0240475  &$18^h56^m51.9^s$  &$-02^\circ40'47.4"$  &16.01  &  \nodata  &       \nodata             &         \nodata      &MISC  &\\
SRc02\_00539  &    &18571441-0216385  &$18^h57^m14.4^s$  &$-02^\circ16'38.8"$  &15.51  &   56.768  &    $0.06263 \pm 0.00003$  &          $0.03 \pm 0.03$  &DSCT  &\\
SRc02\_00589  &c   &18563332-0301561  &$18^h56^m33.3^s$  &$-03^\circ01'56.1"$  &14.08  &  \nodata  &       \nodata             &         \nodata      &MISC  &\\
\enddata
\tablecomments{The flag $c$ denotes stars affected by crowding. Previously known objects are flagged with $k$.
Their IDs from VSX or GCVS can be found in the last column.
$R_B$ is the apparent magnitude in BEST II photometric system.
The Epoch $T_0$ is given in reduced Julian date [rHJD] in respect to {\bf $T=2,454,900.0$}.
It denotes the first minimum in the light curve.
$P$ is the period of the light variaton and $A$ is the amplitude of the variability.
This table is published in its entirety in the electronic edition
of the {\it Astronomical Journal}. A portion is shown here for guidance regarding its form and content.}
\end{deluxetable}

\subsection{Classification}
\label{classification}

The classification of the periodic variable
stars was based on the period, amplitude and shape of their light curve
according to a simplified scheme based on the General Catalog of Variable Stars
\citep[GCVS, ][]{samu09}.

Intrinsic variable stars were sorted into
Delta Scuti (DSCT), $\gamma$ Doradus (GDOR), Cepheid (DCEP), RR Lyrae
(RR) and Beta Cephei (BCEP) types.

Eclipsing binary stars were classified as detached (Algol
type, EA), semi-detached (Beta Lyrae type, EB), or contact (W Ursae Majoris
type, EW) systems. The subtype of one eclipsing binary remains undefined, it is marked as E.

Variables having sinusoidal-like light curves are classified as
ellipsoidal variables (ELL). Stars with light curves that exhibit
features of starspots are marked as spotted stars (SP).
The light variation in these objects is caused by stellar rotation.

Stars that vary on time scales longer than the observational baseline
are classified as probably long periodic (LPV).
Stars with periodic variations and unstable light curve were marked as
$\alpha^2$ Canum Venaticorum (ACV) stars.
Non-periodic variables were classified as miscellaneous (MISC).
Due to the insufficient coverage of the epochs of the observations,
numerous periodic variable stars can be hidden in this category.
In the case of questionable light curves we marked as
mixed types (EW/DSCT, ELL/SP).

An overview of the classification result is given in Table \ref{classification_table}.

\begin{table}[ht]
\begin{center}
\caption{Variable stars in and around the CoRoT SRc02 field. The number of newly detected variables
are in brackets.}
\label{classification_table}
\begin{tabular}{lc}
\noalign{\smallskip}
\hline
\hline
\noalign{\smallskip}
Type & N \\
\noalign{\smallskip}
\hline
\noalign{\smallskip}
\noalign{\smallskip}
\multicolumn{2}{l}{Intrinsic variables}\\
\noalign{\smallskip}
DSCT    &  83  (83)\\
GDOR    &  13  (13)\\
DCEP    &  13  (13)\\
RR      &  24  (24)\\
BCEP    &   2   (2)\\
\noalign{\smallskip}
\hline
\noalign{\smallskip}
\noalign{\smallskip}
\multicolumn{2}{l}{Extrinsic variables}\\
\noalign{\smallskip}
EA      & 125 (124)\\
EW      &  77  (76)\\
EB      &  38  (38)\\
E       &   1   (1)\\
ELL     &  44  (44)\\
ELL/SP  &  16  (16)\\
SP      &  44  (44)\\
ACV     &  37  (37)\\
\noalign{\smallskip}
\hline
\noalign{\smallskip}
\noalign{\smallskip}
\multicolumn{2}{l}{Variables with unclear physical process}\\
\noalign{\smallskip}
LPV      & 453 (426)\\
EW/DSCT &  38  (38)\\
MISC    & 838 (837)\\
\noalign{\smallskip}
\hline
\noalign{\smallskip}
All     & 1846 (1816)\\
\noalign{\smallskip}
\hline
\hline
\end{tabular}
\end{center}
\end{table}

\section{Firsts results}
\label{firstresults}

\subsection{Known variables}
\label{known_var}

The stars observed with BEST II are cross-checked with the variable star
index\footnote{http://www.aavso.org/vsx/} (VSX) of the American Association
of Variable Star Observers (AAVSO) and with the General Catalogue of Variable Stars
\citep[GCVS,][]{samu09}. In the variable star catalogue (Table \ref{tab:varcat})
the previously known stars are marked with a flag 'k'.

Within the observed field there are 30 previously known variable stars.
We could confirm the variability for all of these stars. 27 variables
belong to the long period variables (LPV) type.

NSVS 13967794 was marked as an irregular variable, which is consistent to MISC in our data set.
V914 Aql is an Algol type eclipsing binary with an orbital period of $3.33722 \pm 0.00001$ days
and with quite different primary and secondary eclipse depths. This indicates a large
difference between the masses of the components.

\subsection{Newly detected variables}
\label{new_var}

The most populous group is the long periodic variables with 453 objects. This is distinctive among
the fields observed by BEST II. Another large group is the class of eclipsing binaries
(EA, EB, EW and E type stars) with a total of 241 members. A more detailed study
of these binaries is discussed in Section \ref{binaries}.

The large number of variables in the MISC group is partially due to the insufficient epoch
distribution, which does not allow us to classify more precisely.

There are many interesting variable stars in our dataset.
Here we present only two of these objects. They are shown in Fig. \ref{fig:lc_sample}.

\subsubsection{$\rm{SRc02\_30243}$}

SRc02\_30243 is an RR Lyrae with a pulsation period of 0.46 days
showing the Blazhko effect \citep{blaz07}.
This effect is a periodic modulation of the light variation in the range of tens of days.
The origin of this modulation is still a puzzle after more than a hundred years after
its discovery.

\subsubsection{$\rm{SRc02\_78984}$}

This star has a similar light curve to the new class of variable young stellar object
found by \citet{rodrigez-ledesma12} in the Orion Nebula Cluster,
\citet{klagyivik13} in the young open cluster NGC 2264 and \citet{pawlak13}
in the Small Magellanic Cloud. This kind of
light curve shape can be reproduced with a hot spot either on the star
or on an accretion disk around it.

\section{Eclipsing binaries}
\label{binaries}

In a recent paper we developed a simple model to calculate the fraction
of observable Algol type eclipsing binaries in a random field \citep{klagyivik13}.
On that observational run this model results a fraction of $0.104 \pm 0.004 \%$.
In the SRc02 field we identified 125 Algols among the total of 86,944 stars,
which means $0.14 \pm 0.02 \%$. This is in fairly good agreement - within 2$\sigma$ - with the model.

\subsection{Notes on selected eclipsing binaries}
\label{models}

We study individual eclipsing binaries in more detail by modeling their light curve
with our own code described by \citet{csizmadia09}.
All details of the modeling are the same as in \citet{klagyivik13}.
The selection criteria are the brightness and the amplitude, for
bright binaries with deep eclipses are easier to model correctly.

\begin{figure*}
\centering
  \includegraphics[width=0.32\textwidth]{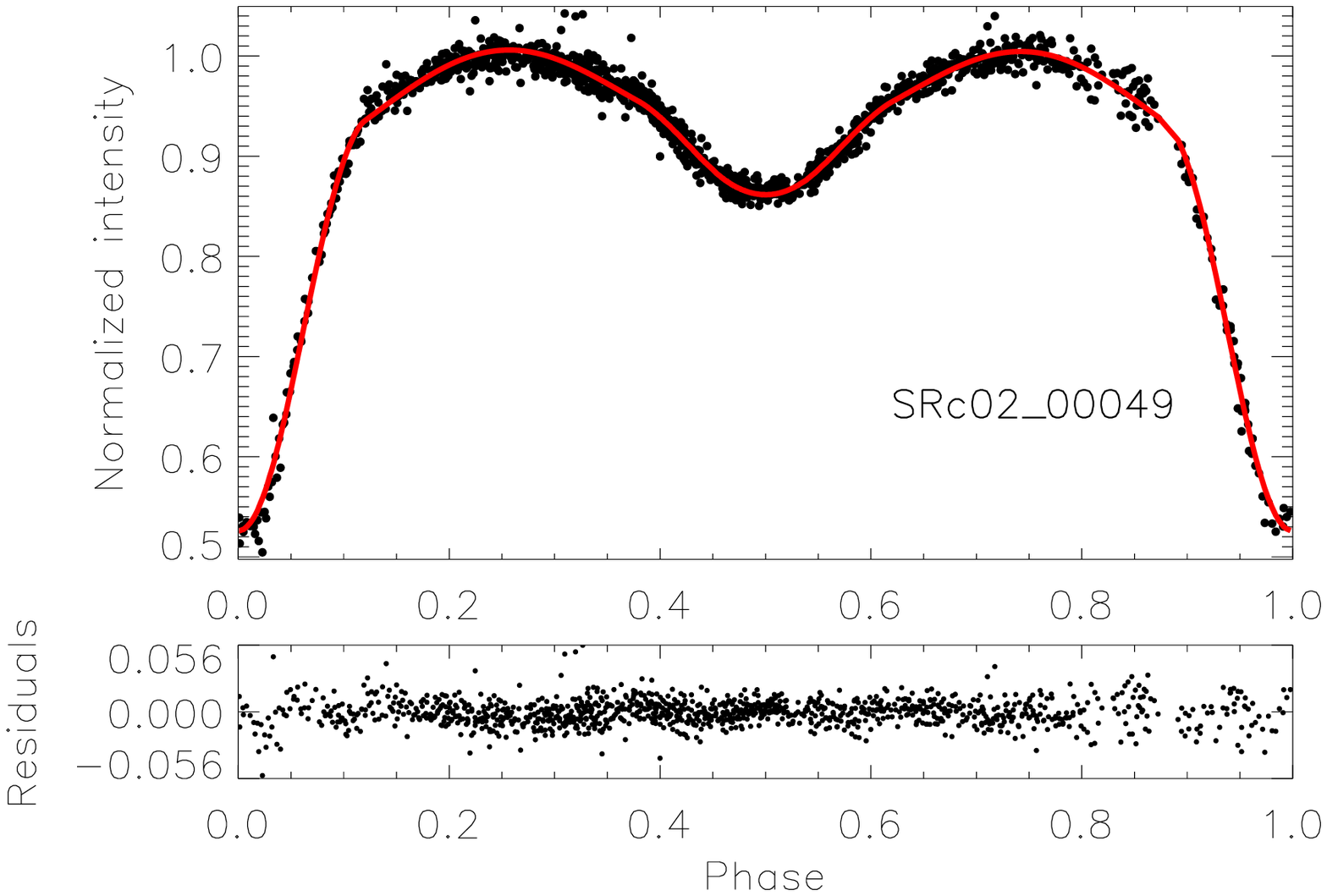}
  \includegraphics[width=0.32\textwidth]{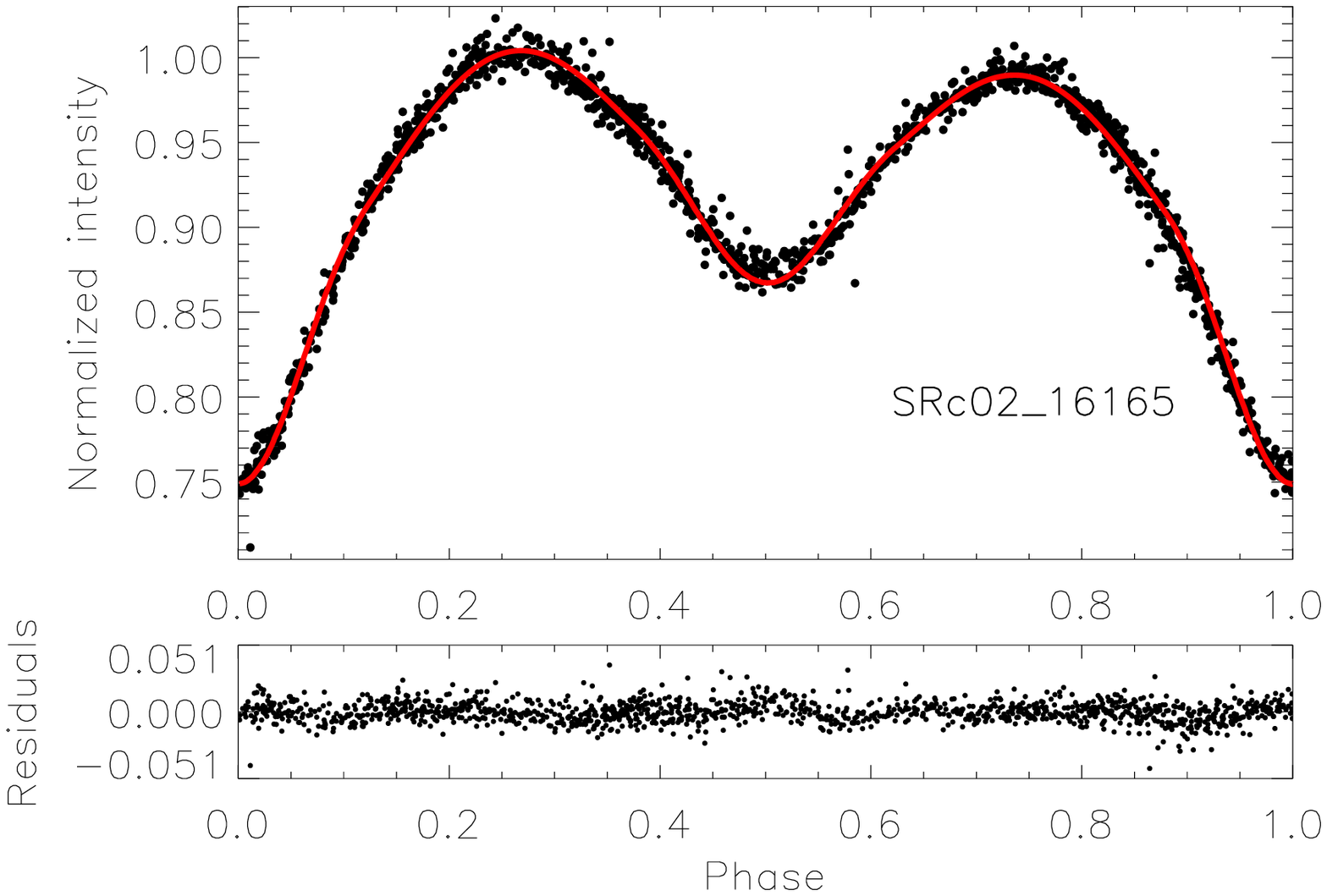}
  \includegraphics[width=0.32\textwidth]{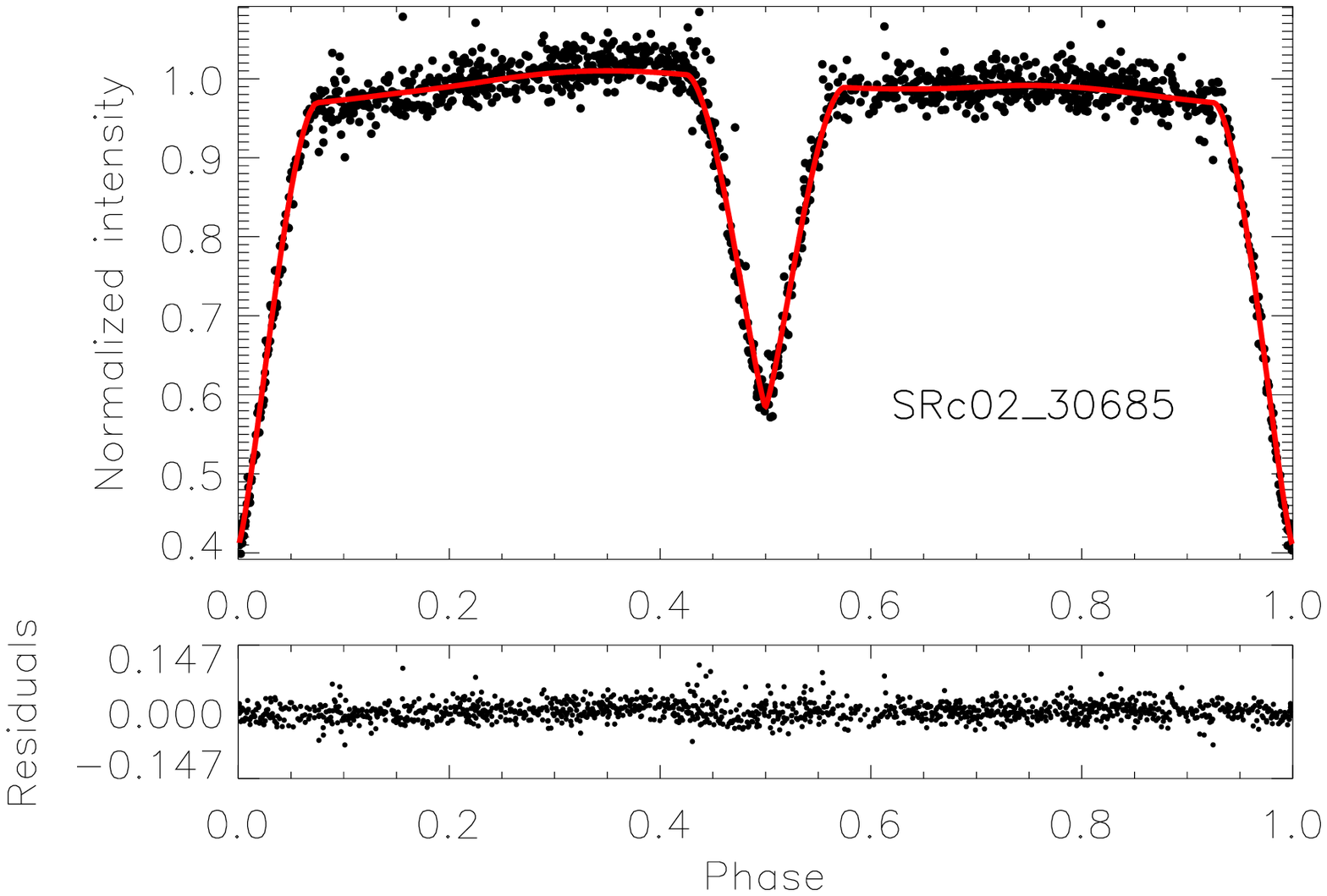}
  \includegraphics[width=0.32\textwidth]{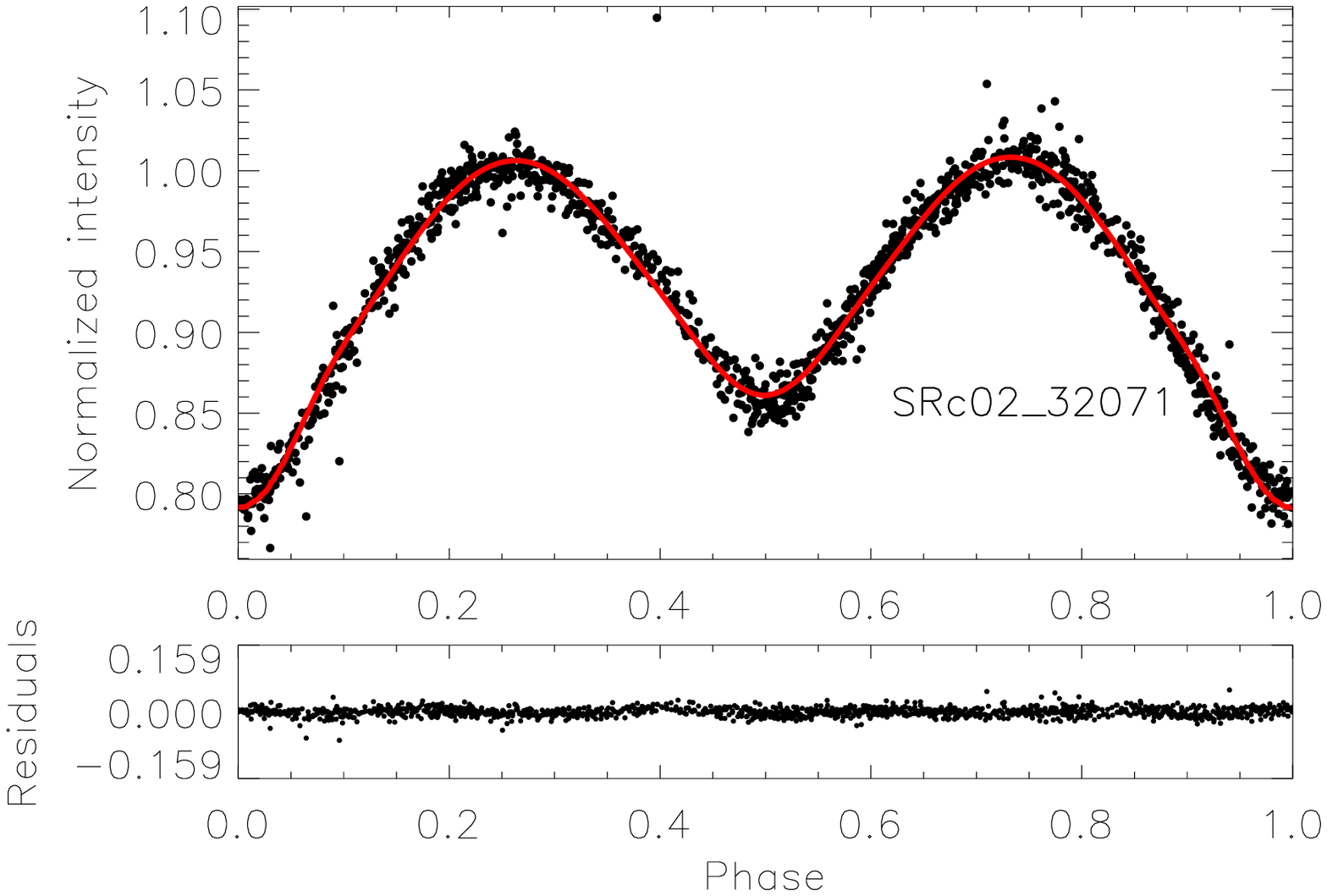}
  \includegraphics[width=0.32\textwidth]{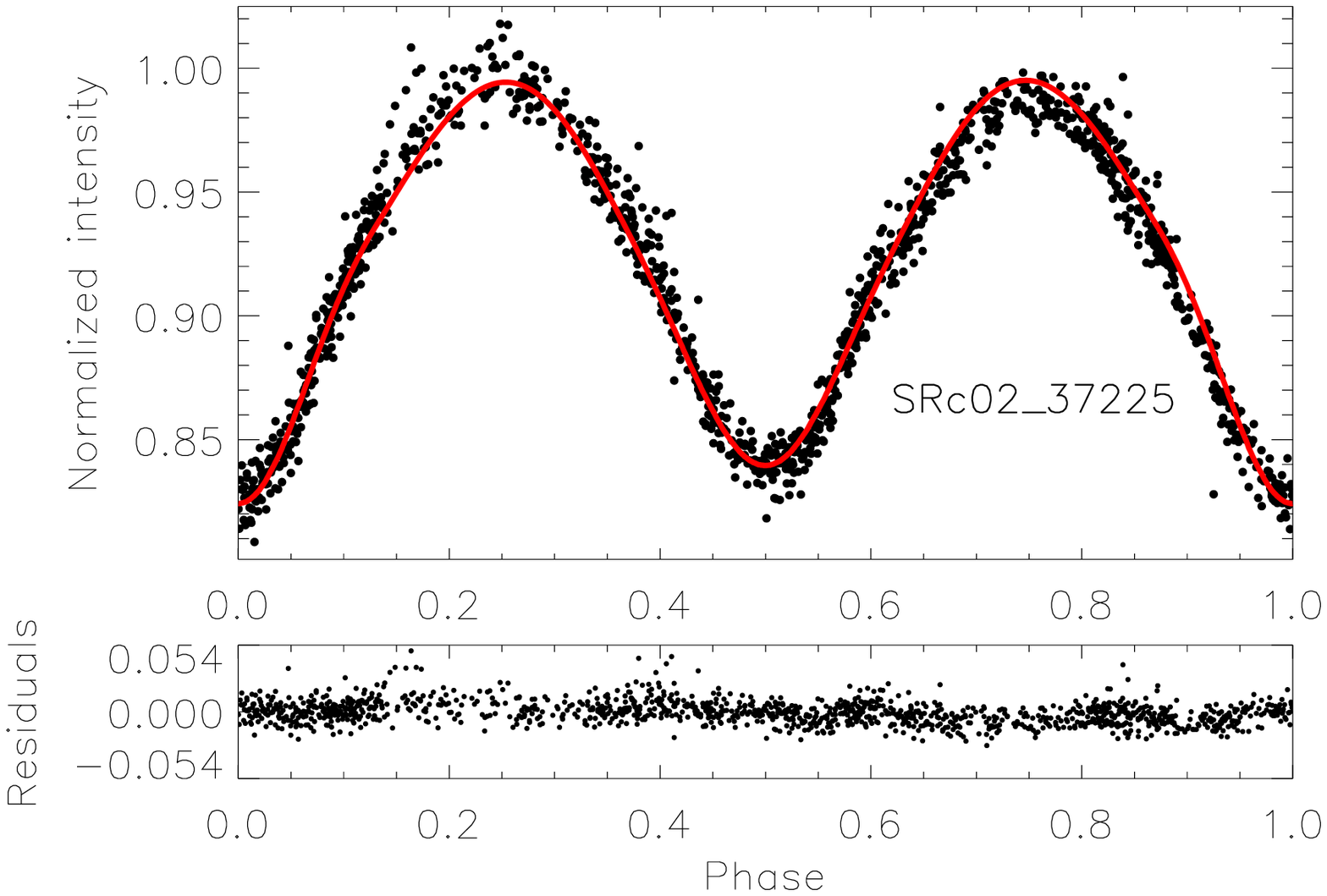}
  \includegraphics[width=0.32\textwidth]{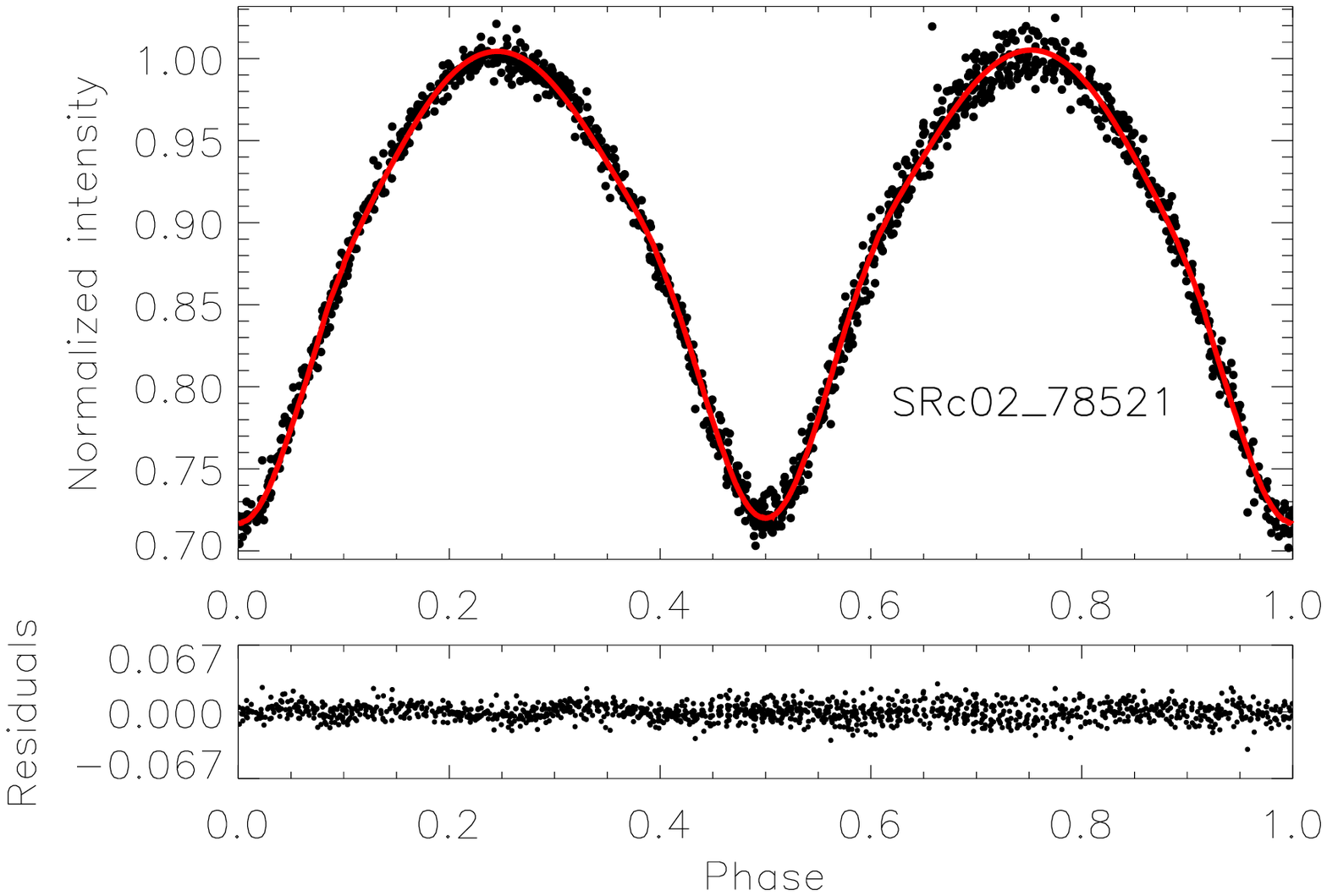}
  \includegraphics[width=0.32\textwidth]{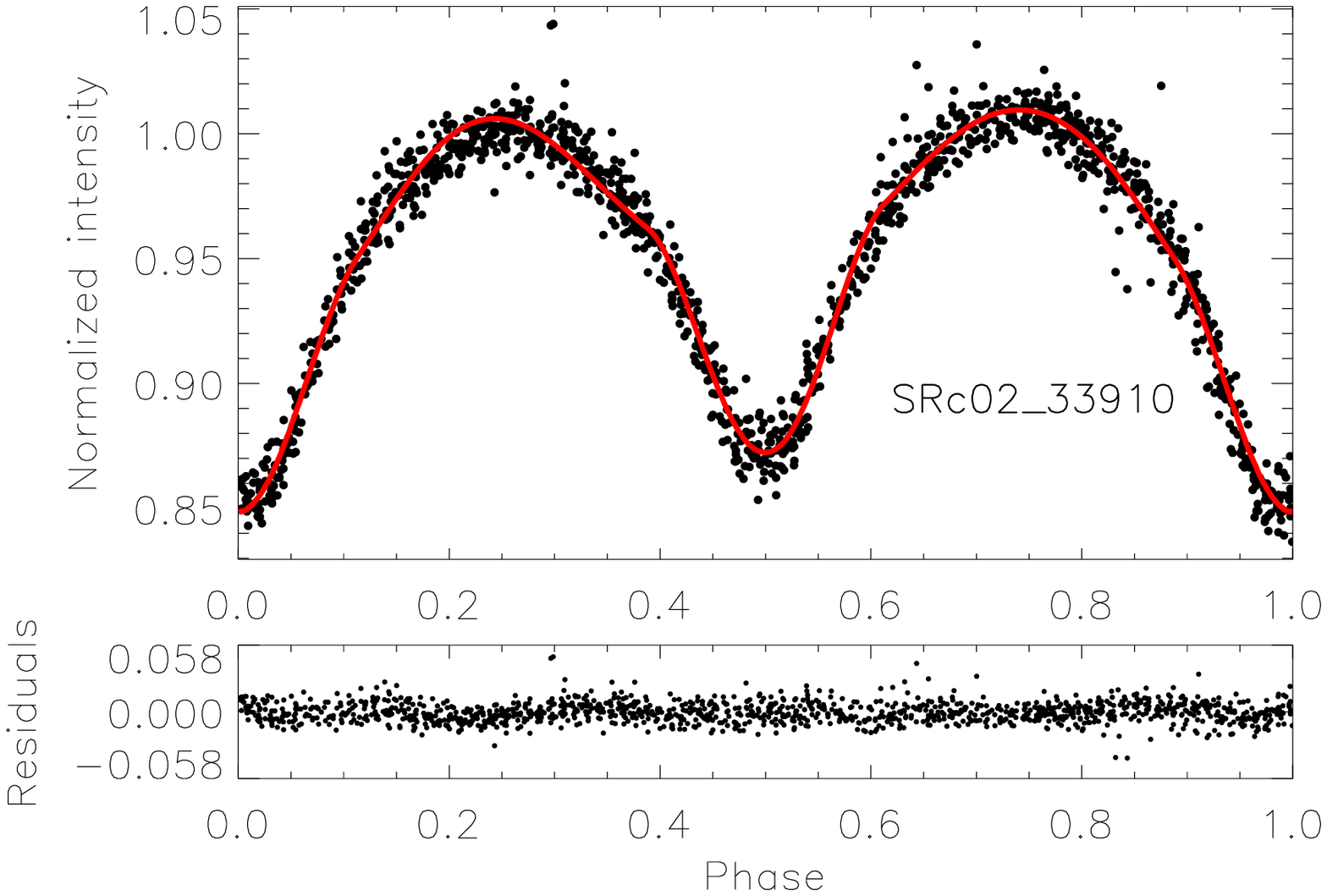}
  \includegraphics[width=0.32\textwidth]{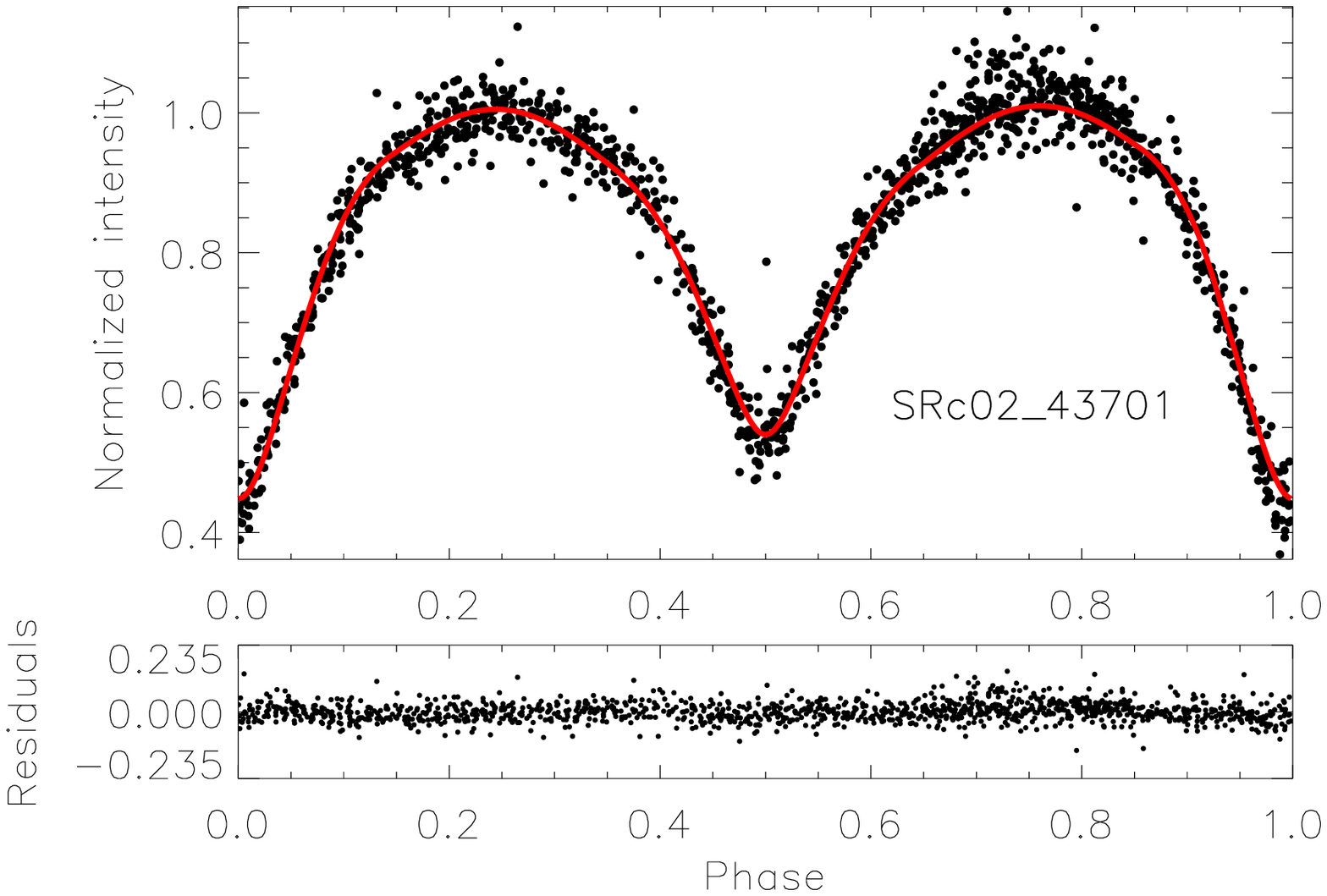}
  \includegraphics[width=0.32\textwidth]{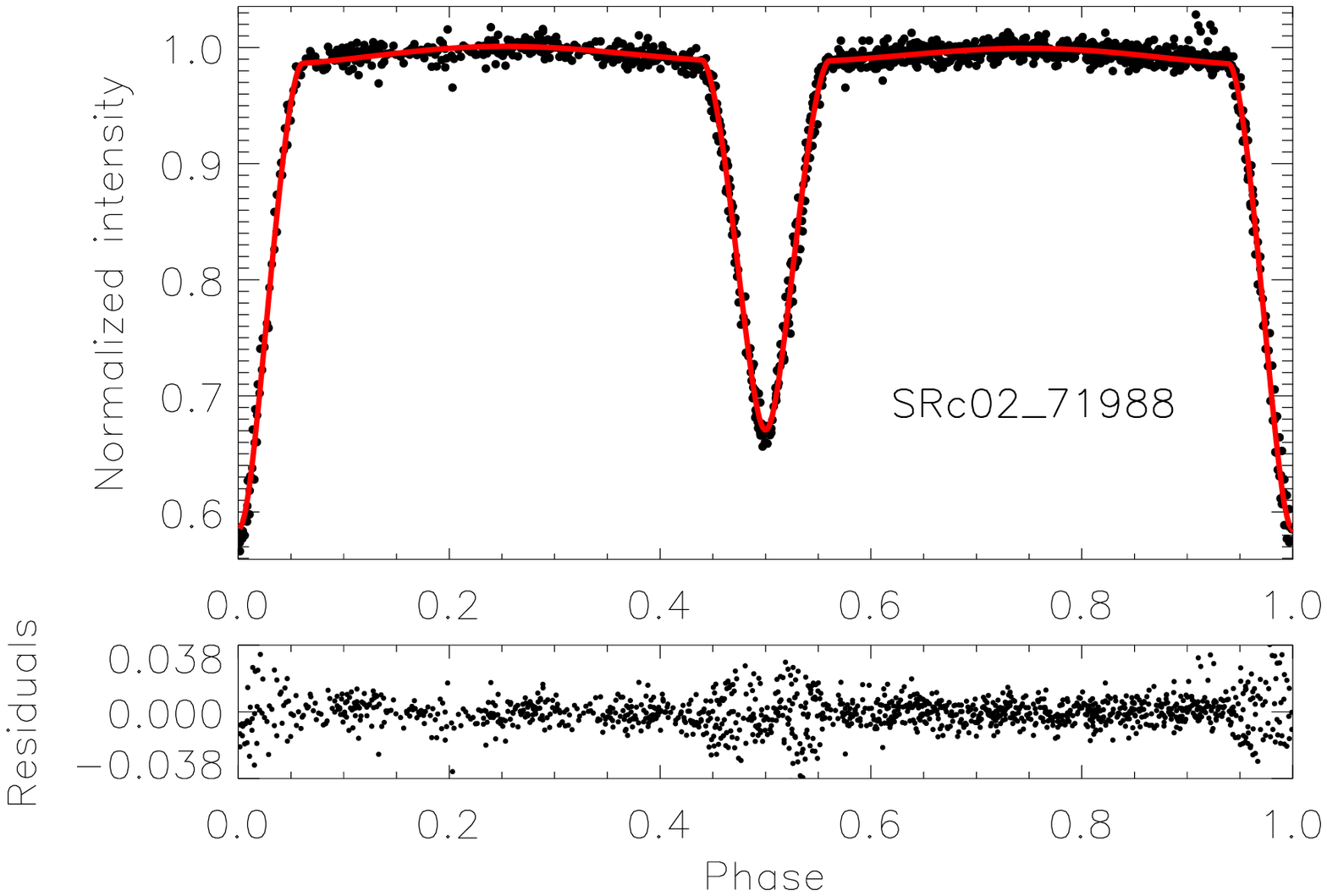}
\caption{Results of the light curve modeling of the selected eclipsing binaries.
The points represent the raw observations, while the solid lines (colored red in the online version)
are the fits. The lower panels show the residuals.}
\label{fig:lcs_modelled}
\end{figure*}

The effective temperature of the primaries are calculated using the 2MASS $J-K$
data. Note that this method can introduce systematic errors due to unknown reddening.
We use the $R$-band linear bolometric and quadratic limb darkening coefficients
published by \citet{vanhamme93} and \citet{claret11}, respectively, since this is the
closest filter to our filterless observations.
Since all stars are cooler than $6000$\,K we fix the gravity darkening exponents
to $g=0.32$ and the albedos to $A=0.5$ \citep[cf.][]{lucy67,rucinski69}.

The free parameters are: mass ratio, inclination, fill-out factors of the two components,
effective surface temperature of the secondary star, epoch, and height of the maximum brightness
refining the normalization of the light curve.
We add a stellar spot to one or both of the components if needed.
We try all the possible combinations up to 2 spots in total (no spot, one spot on the
primary star, one spot on the secondary component, etc.).
When the fitted mass ratio ($q = m_2/m_1$) is higher than 1.0 we reverse the components to keep the primary
component the more massive.
The temperature of a spot is described as the temperature ratio of the spot and the star
(temperature factor = $T_{\rm spot}/T_{\rm star}$). The fits with the smallest
$\chi^2$ value are accepted and are summarized in Table \ref{tab:model_parameters}
and Figure \ref{fig:lcs_modelled}. The errors in Table \ref{tab:model_parameters}
include only modeling uncertainties but no systematic errors of the input parameters.
The individual systems are discussed below.

\begin{table*}
\centering
\caption{Fitted parameters of the modeled binary systems. The temperature factor of the
spots means $T_{spot}/T_{star}$.}
\label{tab:model_parameters}
\small
\begin{tabular*}{0.75\textwidth}{lcccccccccc}
\noalign{\smallskip}
\tableline
\tableline
BEST ID                              & SRc02\_00049        & SRc02\_16165          & SRc02\_30685        \\
\tableline
Orbital period (days)                & $0.9233 \pm 0.0003$ & $0.39607 \pm 0.00006$ & $0.7416 \pm 0.0005$ \\
Mass ratio $q$                       & $0.21\pm0.03$       & $0.21\pm0.02$         & $0.65\pm0.06$       \\
Inclination $i$ ($^{\circ}$)         & $68.2\pm2.0   $     & $59.0\pm1.0   $       & $89.9\pm0.5   $ 	 \\
Fill-out factor $f_1$                & $-0.91\pm0.52 $     & $-0.55\pm0.30 $       & $-5.29\pm0.37 $ 	 \\
Fill-out factor $f_2$                & $-0.54\pm0.12 $     & $ 0.00\pm0.10 $       & $-2.39\pm0.16 $ 	 \\
Temperature $T_1$ (K)                & $3743\pm234   $     & $3558\pm48    $       & $3720$ (fixed)  	 \\
Temperature $T_2$ (K)                & $5029$ (fixed)      & $4893$ (fixed)        & $3772\pm35   $  	 \\
{\it Spot}                           &                     &		           &		     	 \\
{\it On which star?}                 & primary             & secondary             &  primary	     	 \\
Colatitude $\phi_{1}$ ($^{\circ}$)   & $179\pm19  $        & $111\pm64  $          & $136\pm34   $   	 \\
Longitude $\lambda_{1}$ ($^{\circ}$) &  $127\pm142$        & $111\pm56 $           & $200\pm4	 $   	 \\
Diameter $d_{1}$ ($^{\circ}$)        &  $65\pm34    $      & $19\pm10  $           & $9.5\pm3.6  $   	 \\
Temperature factor                   & $0.66\pm0.59  $     & $0.84\pm0.25  $       & $1.26\pm0.07 $      \\
\tableline
$\chi^2$                             & $0.757        $     & $0.647        $       & $0.779        $     \\
\tableline
\tableline
BEST ID                              & SRc02\_32071        & SRc02\_37225        & SRc02\_78521 	 \\
\tableline
Orbital period (days)                & $0.7377 \pm 0.0003$ & $1.3094 \pm 0.0007$ & $0.38834 \pm 0.00004$ \\
Mass ratio $q$                       & $0.22\pm0.02 $      & $0.202\pm0.006$     & $0.52\pm0.08$	 \\
Inclination $i$ ($^{\circ}$)         & $55.6\pm0.7   $     & $55.7\pm0.5   $     & $66.2\pm0.3$ 	 \\
Fill-out factor $f_1$                & $0.11\pm0.01 $      & $0.00\pm0.21$       & $0.04\pm0.01$	 \\
Fill-out factor $f_2$                & $0.11\pm0.01$       & $0.00\pm0.21$       & $0.04\pm0.01$	 \\
Temperature $T_1$ (K)                & $3877\pm39$         & $4115\pm52 $    	 & $4664$ (fixed)	 \\
Temperature $T_2$ (K)                & $4790$ (fixed)      & $3968$ (fixed)  	 & $4582\pm55$  	 \\
{\it Spot}                           &                     &		         &			 \\
{\it On which star?}                 & secondary           & --  		 & secondary		 \\
Colatitude $\phi$ ($^{\circ}$)       & $162\pm18   $       & --     		 & $159\pm42$		 \\
Longitude $\lambda$ ($^{\circ}$)     & $262\pm109  $       & --     		 & $46\pm38$		 \\
Diameter $d$ ($^{\circ}$)            & $42\pm17    $       & --     		 & $15\pm12$		 \\
Temperature factor                   & $0.14\pm0.40 $      & --                  & $1.73\pm0.28$	 \\
\tableline
$\chi^2$                             & $0.919        $     & $0.525        $     & $1.269$		 \\
\tableline
\tableline
BEST ID                              & SRc02\_33910	     & SRc02\_43701          & SRc02\_71988      \\
\tableline
Orbital period (days)                & $0.28172 \pm 0.00004$ & $0.36663 \pm 0.00007$ & $1.371 \pm 0.002$ \\
Mass ratio $q$                       & $0.19\pm0.03$	     & $0.72\pm0.10 $        & $0.64\pm0.28$     \\
Inclination $i$ ($^{\circ}$)         & $62.5\pm0.6  $	     & $83.5\pm0.6   $       & $84.5\pm0.4  $    \\
Fill-out factor $f_1$                & $-0.72\pm0.08 $       & $-0.11\pm0.05 $       & $-6.8\pm2.1 $     \\
Fill-out factor $f_2$                & $-0.80\pm0.14 $       & $-0.07\pm0.04$        & $-4.3\pm0.3 $     \\
Temperature $T_1$ (K)                & $4172$ (fixed)	     & $4128\pm30    $       & $4734\pm52    $   \\
Temperature $T_2$ (K)                & $3576\pm18    $       & $3847$ (fixed)        & $5055$ (fixed)    \\
{\it Spot}                           &  		     &                       &		         \\
{\it On which star?}                 & primary  	     & secondary             & secondary         \\
Colatitude $\phi$ ($^{\circ}$)       & $148\pm48   $	     & $104\pm62  $          & $39\pm86	   $     \\
Longitude $\lambda$ ($^{\circ}$)     & $126\pm141  $	     & $18\pm50   $          & $100\pm102 $      \\
Diameter $d$ ($^{\circ}$)            & $11\pm9     $	     & $7.6\pm4.1  $         & $4.9\pm2.2  $     \\
Temperature factor                   & $2.0\pm0.5  $	     & $1.14\pm0.12  $       & $0.83\pm0.25 $    \\
\tableline
$\chi^2$                             & $1.290	     $       & $0.583        $       & $1.286        $   \\
\tableline
\end{tabular*}
\end{table*}

\subsection{$\rm{SRc02\_00049}$}
\label{00049}

This system consists of two very ellipsoidal stars, as one can calculate
from the fill-out factors and the mass ratio.

The primary star has a polar fractional radius (defined as $R/a$, where $R$ is the
actual stellar radius and $a$ is the semi-major axis of the orbit) of 0.457 $\pm$ 0.004,
and the equatorial stellar fractional radii are 0.505 $\pm$ 0.005
in the direction of the other star, 0.536 $\pm$ 0.005 to the anti-companion direction, and
0.489 $\pm$ 0.004 into the direction of motion (these are called polar, point, back and side
radii according to the binary star terminology).
The secondary star has fractional radii of $r_{pole} = 0.219 \pm 0.002$, $r_{point} = 0.263 \pm 0.003$,
$r_{back} = 0.248 \pm 0.002$ and $r_{side} = 0.226 \pm 0.002$.

The polar oblateness, defined as $1-r_{pole}/r_{side}$, is nearly 7\% for the primary
and 3\% for the secondary. Although the two stars are quite close to each other,
they are far from Roche-lobe filling. The primary has an average radius of 74\%
of its own Roche-lobe size, while the secondary is only 67\% of its own
Roche-lobe size. Therefore it seems they have not undergone mass-transfer yet,
and hence they have their original masses. The light curve modeling was
able to reproduce the observed flux variations with synchronous rotation, so
they are fast rotators, which means that the stellar shape is deformed by
strong tidal rotation. These, and the fast rotation make this and similar objects good targets
to determine and to study the relationship between internal stellar sturcture
(and $J_2$ parameter) and stellar shapes under strong rotational and tidal
interactions.

Although the light curve fit required a spot, the temperature factor of the spot
is poorly determined and it allows also an unspotted star.

\subsection{$\rm{SRc02\_16165}$}
\label{16165}

This object is a semi-detached binary where the secondary fills out the
Roche-lobe. The O'Connell-effect is clearly visible, the two maxima differ from each
other by 0.01 magnitude and it seems to be stable during the observational
window. This was reproduced by a dark spot on the secondary. One can expect
period variations during the mass transfer, and observations of this variation
on longer time-scale would be important to establish which phase of the mass
transfer occurs in this system. Its brightness makes it a favourable target for
further spectroscopic and photometric investigation.

\subsection{$\rm{SRc02\_30685}$}
\label{30685}

SRc02\_30685 is a high-amplitude
system with a considerable reflexion effect and light curve distortion. From the end of the
primary minimum to the beginning of the secondary minimum, the flux level
increases by 0.04 magnitude due to reflection. However,
from the secondary to the primary minimum the flux level is close to constant
what we explain by a spot. Actually, this bright spot is necessary on the
primary to explain the observed out-of-eclipse variations. Such spots are
common in binaries, but less frequent than the dark spots.

\subsection{$\rm{SRc02\_32071, 37225, 78521}$}
\label{32071}

These three systems are so-called contact or overcontact binary systems, where
the two stars are so close to each other that both components fill out their own
Roche-lobe, and thus they are geometrically in contact,
or -- in the case of overcontact systems -- the two stellar cores have a common convective
envelope.

SRc02\_32071 is a low-inclination ($i = 56^{\circ}$) overcontact binary system with 11\%
over-filling factor. It belongs to the W-subtype of overcontact binaries according
to the definition of Binnendijk (1965), since the secondary object is smaller
($r_{avg} = 0.253$) and warmer ($T_{eff} = 4790\,K$) than the primary
($r_{avg} = 0.525$ and $T_{eff} = 3877\,K$).
The secondary maximum is slightly at higher flux level (by 0.01 mag) than the primary maximum,
indication a negative O'Connell-effect. The O'Connell-effect was observed in several
other, well-studied systems, too, but only the minority, 24\% of the systems (19 out of
78) showed such a negative value in the list of \citet{maceroni96}.

SRc02\_37225 is an unspotted contact binary. It seems to be in the state of exact Roche-lobe filling.
According to theoretical expectations, exact Roche-lobe filling is quite unlikely because it
lasts for an extreme short time-interval. It brakes or it evolves to
overcontact. More precise photometry is needed to confirm that we have a
fill-out factor of zero, so a rare type of contact binaries have been found, or
we have a very low fill-out factor in this system. If it is in exact contact,
then it can be an important test object for theoretical studies.

SRc02\_78521 is a common A-subtype overcontact binary with well-determined fill-out
factor ($f_1 = f_2 = 0.04 \pm 0.01$). In case of contact binaries, long time-scale
magnetic cycles (10-20 years, e.g. Borkovits et al. 2005), spot-induced Eclipse
Timing Variations \citep{tran13}, effect of mass transfer, third bodies and
irregular variations occur \citep{borkovits05, tran13, quian01a, quian01b, quian03, nelson14}.
Since this target is bright for even small
telescopes, it is a good target for such period variation studies.

\subsection{$\rm{SRc02\_33910, 43701, 71988}$}
\label{33910}

These three systems are detached Algol-type eclipsing binaries. The light curves
of SRc02\_33910 and SRc02\_43701 show some similarities to each other, but SRc02\_43701 is almost a
near-contact binary while SRc02\_33910 is a wider system which probably evolves later to a
near-contact system. SRc02\_71988 is a well-detached Algol-type system.

The SRc02\_33910 system requires a new light curve solution based on multi-colour photometry
which helps to establish the exact temperature difference between the
components. It seems to be a short-period, detached system.

SRc02\_43701 has average fractional radii $r_{pri} = 0.397 \pm 0.004$ and $r_{sec} = 0.339 \pm 0.003$.
The sum of the point fractional radii are exactly 0.900 $\pm$ 0.008
(0.478 $\pm$ 0.004 and 0.422 $\pm$ 0.004 for the primary and secondary star, respectively).
The inner Lagrange-point is at a distance of 0.53 $\pm$ 0.01 from the primary's center, while
the primary's point radius is 0.478 $\pm$ 0.004, quite close to the inner Lagrange-point.
That is why this system is a pre-near-contact binary, and most likely it will
later evolve to a contact system.

The residual light curve of SRc02\_71988 has a significantly larger scatter in
the primary and secondary minima than out of transit. This can be due to
surface brightness inhomogeneties (e.g. stellar spots) on the side faced
toward its companion, which is occulted during the eclipses.

\section{Summary}
\label{summary}

We presented a study of the variable stars in the CoRoT field SRc02
observed by BEST II in 32 nights in 2009. We detected 1,846
variable stars out of 86,944 stars in our field of view, out of
which 1,816 are new detections and only 30 were previously known.
Most of them are long period variables (classified as LPV),
or show irregular variability (MISC class).

The total number of eclipsing binaries is 241.
There are 125 Algol-type eclipsing binaries in our database, which
means $0.14\%$ of all stars. This is in good agreement with
our simple model of EAs \citep{klagyivik13}.

We studied the light curves of 9 eclipsing binaries using our own
modeling code. SRc02\_37225 seems to be in exact Roche-lobe filling
and it can be an important test object for theoretical studies,
while SRc02\_43701 is a pre-near-contact system and will probably evolve to a contact system.

A new RR Lyrae showing Blazhko modulation was also found (SRc02\_30243), however, the observational run
was not long enough to determine the period of the Blazhko cycle.

\acknowledgments

Peter Klagyivik acknowledges support from the Hungarian State E\"otv\"os Fellowship
and support by grant AYA2012-39346-C02-02 of the Spanish Ministerio de Econom\'ia y Competividad (MINECO).
Petr Kabath would like to acknowledge funding from the Purkyne Fellowship
programme of the Czech Academy of Sciences.
This project has been partly supported by the Hungarian OTKA Grant K113117.
This publication is supported as a project of the Nordrhein-Westf\"alische Akademie der
Wissenschaften und der K\"unste in the framework of the academy programme
by the Federal Republic of Germany and the state Nordrhein-Westfalen.
This research has made use of the SIMBAD database, operated at CDS, Strasbourg, France.
We also made use of 2MASS, GCVS catalogs, and AAVSO variable star search index.

\end{document}